\documentclass[12pt,psfig,epsf]{article}

\usepackage{amsmath,amsfonts,amssymb,amscd}
\usepackage{latexsym,mathrsfs,amsthm,inputenc,fontenc}
\usepackage{epsf}
\usepackage{graphicx}

\newtheorem{theorem}{Theorem}
\newtheorem{lemma}[theorem]{Lemma}
\newtheorem{proposition}[theorem]{Proposition}
\newtheorem{corollary}[theorem]{Corollary}

\newtheorem{remark}{Remark}

\newcommand{\C}{\ensuremath{\mathbb{C}}}
\newcommand{\N}{\ensuremath{\mathbb{N}}}
\newcommand{\Z}{\ensuremath{\mathbb{Z}}}

\newcommand{\R}{\ensuremath{\mathbb{R}}}
\newcommand{\bR}{\overline{\R}}
\newcommand{\bRp}{\overline{\R_+}}
\renewcommand{\S}{\mathbb{S}}

\def\F{\mathcal F}
\def\B{\mathcal{B}}
\def\H{\mathscr H}
\def\Hi{\mathcal H}
\def\Pv{\mathrm{Pv}}
\def\e{e}
\def\o{o}
\def\d{\mathrm{d}}
\def\f{\mathfrak{f}}
\def\SS{\mathcal S}
\def\T{\mathcal T}
\def\F{\mathcal F}
\def\E{\mathcal{E}}
\def\CC{\mathcal C}
\def\J{\mathcal{J}}
\def\K{\mathcal{K}}
\newcommand{\tr}{\mbox{tr}}
\newcommand{\Tr}{\mbox{Tr}}
\renewcommand{\d}{{\mathrm d}}
\newcommand{\Om}{\Omega}

\newcommand{\ind}{\mbox{\rm\ ind}}
\renewcommand{\index}{\mbox{\rm\ index}}

\begin{document}

\title{On the structure of the wave operators in one dimensional
potential scattering}

\author{Johannes Kellendonk\,~and\,~Serge Richard}

  \date{\small
    \begin{quote}
      \emph{
    \begin{itemize}
    \item[]
Universit\'e de Lyon, Universit\'e Lyon 1, Institut Camille Jordan, CNRS UMR 5208, 43 boulevard du 11 novembre 1918, F-69622 Villeurbanne cedex,
France
    \item[]
      \emph{E-mails\:\!:}
      kellendonk@math.univ-lyon1.fr\,~and\,~richard@math.univ-lyon1.fr
    \end{itemize}
      }
    \end{quote}
\today
  }
\maketitle

\begin{abstract}
In the framework of one dimensional potential scattering we prove
that, modulo a compact term, the wave operators
can be written in terms of a universal operator
and of the scattering operator.
The universal operator is related to the one dimensional Hilbert
transform and can be expressed as a function of the generator of dilations.
As a consequence, we show how Levinson's theorem can be rewritten
as an index theorem, and obtain the asymptotic behaviour of
the wave operators at high and low energy and at large and small scale.
\end{abstract}

\section{Introduction}

In recent work we proposed a topological approach to
Levinson's theorem, first
in three dimensions in the generic case, {\it i.e.}~in absence of $0$-energy
resonance \cite{KR2}, and second for one dimensional
scattering systems including also the exceptional cases \cite{KR3}.
One of the key features of this new approach is the use of extensions of
$C^*$-algebras and their associated index map in $K$-theory.
The basic hypothesis we had to make was that
the wave operator $\Om_-$ belongs to a certain $C^*$-algebra.
In this article
we prove the crucial hypothesis for one dimensional
scattering systems with sufficiently fast vanishing potentials $V$.
In fact, we observe an even stronger result, namely that
\begin{equation}\label{newformula}
\Omega_-\big(-\Delta+V,-\Delta\big)  = 1+ \hbox{$\frac{1}{2}$} \big(1 - R(A)\big)\big(S(-\Delta) -
1\big) + K\ ,
\end{equation}
where $S(-\Delta)$ is the scattering operator,
\begin{equation}\label{eq-R}
R(A)= -\tanh(\pi A) - i (P_\e - P_\o ) \cosh(\pi A)^{-1},
\end{equation}
and $K$ is a compact operator. Here $A$ is the generator of dilations
and $P_\e,P_\o$ are the projections onto the even (symmetric), odd
elements of $\H:=L^2(\R)$, respectively.
A similar formula holds for $\Omega_+$,
{\it cf.}~\eqref{omegaplus},
and the same type of formula, but
even with $K=0$, was found for point interactions in \cite{KR1}.

Note that $R(A)$ is universal
in the sense that it does not depend on the potential.
Furthermore, there is a simple relation between
the Hilbert transform $\Hi$ and the operator $R(A)$. In fact, if
$\sigma:\H\to \H$ denotes the operator of multiplication
with the sign of the variable -- it is thus the natural intertwiner between
even and odd functions on $\R$ -- then
$$ \Hi = i\sigma R(A).$$
It thus follows that the wave operator $\Omega_-$ can be rewritten in
terms of $\Hi$ instead of $R(A)$, which is in accordance with the analysis
of the wave operators performed in \cite{DF,W}.

The structure of the wave operator exhibited in \eqref{newformula}
has various implications of which we will discuss two.
First, the observation that Levinson's
theorem can be formulated as an index theorem, and thus is topological
in nature, and second, that certain strong convergent limits in
scattering theory are in fact norm convergent limits in a restricted sense.

\subsection{Levinson's theorem}
As for the first implication recall that
in one-dimensional potential scattering, a common form of Levinson's
theorem reads :
\begin{equation} \label{eq-lev1}
\hbox{$\frac{1}{2\pi}$}\int_0^\infty \tr
[iS^*(\lambda)S'(\lambda)] \d \lambda = N - \nu
\end{equation}
where $N$ is the number of bound states of $H=-\Delta+V$, which is finite under our assumptions,
$S'$ denotes the derivative of $S$ with respect to $\lambda$ and
$\tr$ is the $2\times 2$ matrix trace. The correction term $\nu$
is $\frac{1}{2}$ or $0$ depending on the existence of a resonance for $H$ at
energy $0$.
Using the above obtained formula for the wave operator $\Om_-$, we will
express \eqref{eq-lev1} as an index theorem.

For that purpose, recall that asymptotic
completeness implies that $\Om_\pm$ are isometries with range projection
$1-{P_p}$ where $P_p$ is the projection onto the bound states.
In particular, $\Om_-$ is a Fredholm operator and
\begin{equation*}
N = \Tr(P_p) = -\index(\Om_-)\ .
\end{equation*}
Our aim is to compute
this index in a way similar to the Krein-Gohberg index theorem. We now outline
our approach which is based on the construction of a
norm-closed algebra $\E$ which contains $\Omega_-$ and
sits in between the algebra of compact
operators $\K(\H)$ and that of bounded operators $\B(\H)$ on $\H$:
$\K(\H)\subset\E\subset \B(\H)$. Recall that $\K(\H)$ forms an
ideal in $\B(\H)$, and that $F\in \B(\H)$ is a Fredholm operator if
it is invertible modulo a compact operator,
that is, its image $q(F)$ in the quotient algebra $\B(\H)/\K(\H)$ is
invertible. Its index is a topological (even homotopy) invariant, namely it is
stable against perturbations of $F$ along continuous paths of Fredholm
operators.
Suppose $F$ belongs to $\E$ and that $\E/\K(\H)$
is isomorphic to $C\big(\S,M_2(\C)\big)$, the algebra of continuous
functions over the circle with values in the $2\times 2$ matrices.
Then, viewing $q(F)$ as such a function we can take pointwise its
determinant to obtain a non-vanishing function over
the circle. The winding number of that latter function
is another topological invariant of $F$.
We denote it by $w\big(q(F)\big)$, and we shall show below that these
two invariants are equal. By applying this to $F=\Omega_-$ one obtains
our topological formulation of Levinson's theorem
\begin{equation}\label{eq-lev21}
w\big(q(\Om_-)\big) = -\Tr(P_p)\, .
\end{equation}
Obviously, the sign depends on the choice of orientation for the
winding number.
We shall see in Section \ref{secwin} how the winding number
$w\big(q(\Om_-)\big)$ is related to the l.h.s.~of \eqref{eq-lev1}
and takes account for the correction term $\nu$.

We note that \eqref{eq-lev21} can be refined: if $P$ is a
projection which commutes with $\Omega_-$, then restricting the
analysis to the Hilbert space $P\H$ results in
$w\big(q(\Omega_- P)\big) = -\Tr(P_pP)$. For example, if $V$ is a symmetric
function,  choosing for $P=P_\e$ or $P=P_\o$
leads to a Levinson's theorem for the even or the odd sector.

Whereas to our knowledge it has not yet been pointed out that
Levinson's theorem is a topological theorem, the theorem in its known
form (proved by analytic means) has been used in the context of the
discussion of various topological quantities,
as for instance the Witten-index in supersymmetric quantum mechanics.
We will comment on this in Section \ref{secwin}.

\subsection{Restricted norm convergence}

Further implications of equation~\eqref{newformula}
concern norm estimates which are similar in spirit to propagation
estimates.
It is well known that the relations
\begin{equation*}
s-\lim_{t \to - \infty} e^{iH_0 t}\;\! e^{-iH t}\;\!\Omega_- = 1, \qquad s-\lim_{t \to  +\infty} e^{iH_0 t}\;\! e^{-iH t}\;\!\Omega_- = S
\end{equation*}
hold, with $H_0=-\Delta$ and where the convergence is in the strong topology, but never in the norm topology.
Defining $\ln(H)$ by functional calculus on the positive part
of the spectrum of $H$, the invariance principle implies also that
\begin{equation}\label{lnf}
s-\lim_{t \to - \infty} e^{i\ln(H_0) t}\;\! e^{-i\ln(H) t}\;\!\Omega_- - 1=0, \qquad s-\lim_{t \to  +\infty} e^{i\ln(H_0) t}\;\! e^{-i\ln(H) t}\;\!\Omega_- - S =0
\end{equation}
again involving the strong but not the norm topology.
Now, a rather surprising corollary of our approach is that
convergence of the above limits holds in the norm topology in a
restricted sense, namely after
multiplying \eqref{lnf} with a suitable spectral projection
of the generator $A$ of dilations,
see Prop.~\ref{propev}.
Similar results also hold for expressions in which $A$ is exchanged
with $\ln(H_0)$, see Prop.~\ref{propdil}. All these estimates are related to the asymptotic behaviour of the wave operators at high and low energy and at large and small scale, as explained in Section \ref{sec5}.

\medskip

\noindent
{\bf Acknowledgment} S.R. thanks S. Nakamura for a two weeks
invitation to Japan where part of the present work was completed. This
stay was made possible thanks to a grant from the Japan Society for
the Promotion of Science.

\section{Potential scattering in one dimension}

In this section we introduce the precise framework for our
investigations. We refer to \cite{AY,DT,Sch} for the proofs of the
following standard results.

Let us denote by $\H$ the Hilbert space $L^2(\R)$, and for any $n \in
\N$, let $\H^n$ denote the usual Sobolev space of order~$n$ on
$\R$. We consider for the operator $H_0$ the Laplacian
operator $-\Delta$ with domain $\H^2$. For the perturbation, we assume
that $V$ is a real function on $\R$ that belongs to $L^1_1(\R)$, where
for any $\rho \in \R_+$, $L^1_\rho(\R)$ is the weighted $L^1$-space
\begin{equation*}
L^1_\rho(\R) = \big\{v :\R \to \C \mid \int_\R \langle x \rangle^\rho
\;\! |v(x)|\;\!\d x < \infty \big\}\ ,
\end{equation*}
with $\langle x \rangle= (1+x^2)^{1/2}$. In that situation,
the quadratic form defined by
\begin{equation*}
\H^1 \ni f \mapsto \int_\R \big(|f'(x)|^2 + V(x)|f(x)|^2\big)\;\!\d x
\end{equation*}
is closed, bounded from below, and defines a unique self-adjoint
operator $H$ in $\H$. The spectrum of this operator consists of an
absolutely continuous part equal to $[0,\infty)$ and of a finite
number of eigenvalues which are all located in
$(-\infty,0)$. Furthermore, the wave operators
\begin{equation*}
\Omega_\pm := s-\lim_{t \to \pm \infty} e^{iHt}\;\! e^{-iH_0 t}
\end{equation*}
exist and are asymptotically complete. It then follows that the
scattering operator $S=\Omega_+^*\;\!\Omega_-$ is unitary. Let us
also mention that in the direct integral representation of $\H$
with respect to $H_0$, the spectral decomposition of $S\cong
\{S(\lambda)\}_{\lambda \geq 0}$ satisfies the following property: The
map
\begin{equation*}
\R_+ \ni \lambda \mapsto S(\lambda)\in M_2(\C)
\end{equation*}
is continuous and has limits at $0$ and at $+\infty$ \cite{AK,DT,K}.

Our analysis of the wave operators is based on their representation
in terms of the generalized eigenfunctions
which we briefly recall. A full derivation can be found in
\cite[Sec.~2]{AY}. For simplicity, we shall restrict ourselves to the
study of $\Omega:=\Omega_-$, the analysis of
$\Omega_+$ being then an easy corollary, see equation \eqref{omegaplus}.

For each $x \in \R$ and $k\in \R^*$, let $\Psi(x,k)$ be the solution
of the Lippmann-Schwinger equation:
\begin{equation}\label{LP}
\Psi(x,k)=e^{ikx}+\hbox{$\frac{1}{2i|k|}$}\int_\R
e^{i|k|\;\!|x-y|}\;\!V(y)\;\! \Psi(y,k)\;\!\d y\ .
\end{equation}
For $k\in \R^*$ fixed, the r.h.s.~has a well defined meaning since the map $y \mapsto \Psi(y,k)$ belongs to $L^\infty(\R)$.
Then the wave operator $\Omega$ is formally given on any $f \in \H$ by
\begin{equation}\label{rate}
[\Omega f](x) = \hbox{$\frac{1}{\sqrt{2\pi}}$}\int_\R \Psi(x,k)\;\! \hat{f}(k)\;\! \d k .
\end{equation}
The notation $\hat{f} = \F[f]$ is used for the Fourier transform of $f$ defined on any $f \in L^2(\R)\cap L^1(\R)$ by $\hat{f}(k) = \frac{1}{\sqrt{2\pi}} \int_\R e^{-ikx} f(x)\; \!\d x$.

\section{A new formula for the wave operators}\label{secT}

Starting from the Lipmann-Schwinger equation \eqref{LP} for $x \in \R^*$
and $k \in \R^*$, let us deduce \eqref{newformula} for the wave operator
$\Omega$. The notation $\hat{x}$ is used for the unit vector
$\hbox{$\frac{x}{|x|}$}$. One has:
\begin{eqnarray*}
\Psi(x,k) - e^{ikx} & = & \hbox{$\frac{1}{2i|k|}$}\int_\R
e^{i|k|\;\!|x-y|}\;\!V(y)\;\! \Psi(y,k)\;\!\d y \\
& = & \hbox{$\frac{1}{2i|k|}$}\int_\R
\Big[ e^{i|k|\;\!(|x|-\hat{x}y)} + \Big(
e^{i|k|\;\!|x-y|}-e^{i|k|\;\!(|x|-\hat{x}y)}\Big)\Big]
\;\!V(y)\;\! \Psi(y,k)\;\!\d y \\
& = & e^{i|k|\;\!|x|} \;\!\f(k^2,\hat{k},\hat{x}) + K(x,k) \ ,
\end{eqnarray*}
where
\begin{eqnarray*}
\f(k^2,\hat{k},\hat{x}) & := &
\hbox{$\frac{1}{2i|k|}$} \int_\R
e^{-i|k|\hat{x}y}\;\!V(y)\;\! \Psi(y,k)\;\!\d y \\
K(x,k) &:= & \hbox{$\frac{1}{2i|k|}$}\int_\R \Big(
e^{i|k|\;\!|x-y|}-e^{i|k|\;\!(|x|-\hat{x}y)}\Big)
\;\!V(y)\;\! \Psi(y,k)\;\!\d y\ .
\end{eqnarray*}
We suspect that the expression $\f(k^2,\hat{k},\hat{x})$ is equal to the scattering amplitude for any $V \in L^1_1(\R)$, but have not been able to locate such a result in the literature. However, it is proved in \cite{A,N} that such an equality holds if the potential is slightly more regular, and in particular for $V \in L^1_2(\R)$ which is assumed in the sequel. It then follows from \eqref{rate} and from the relation between the scattering amplitude
and the scattering operator that for any $f$ belonging to the Schwartz space $\SS(\R)$:
\begin{eqnarray*}
[(\Omega-1) f](x)
& = & \hbox{$\frac{1}{\sqrt{2\pi}}$}\int_\R
e^{i|k|\;\!|x|}\;\!\f(k^2,\hat{k},\hat{x})\;\!\hat{f}(k)\;\!\d k
+ \hbox{$\frac{1}{\sqrt{2\pi}}$}\int_\R K(x,k)\;\!\hat{f}(k) \;\!\d k \\
& = & \hbox{$\frac{1}{\sqrt{2\pi}}$}\int_{\R_+} e^{i\kappa |x|}\;\!
\big[(S(\kappa^2)-1)\hat{f}\big](\kappa \hat{x})\;\!\d \kappa
+ [Kf](x)\\
& = & \big[\big(\T\big(S(-\Delta)-1\big)+K\big)f\big](x)
\end{eqnarray*}
where
\begin{equation}\nonumber
[\T f](x) :=  \hbox{$\frac{1}{\sqrt{2\pi}}$}\int_{\R_+}
e^{i\kappa|x|}\;\!\hat{f}(\kappa \hat{x})\;\!\d \kappa
\end{equation}
and
\begin{equation*}
[Kf](x):=\hbox{$\frac{1}{\sqrt{2\pi}}$}\int_\R K(x,k)\;\!\hat{f}(k) \;\!\d k\ .
\end{equation*}
We will show below that for $V \in L^1_\rho(\R)$ with $\rho > \frac{5}{2}$, the kernel $K(\cdot, \cdot)$ belongs to $L^2(\R\times\R)$, and thus the operator $K$ is compact.

Our next aim is to rewrite the operator $\T$ as a function of the dilation
operator $A$. But first we relate it to the Hilbert transform.
Recall that the Hilbert transform on $\R$ is defined on any $f \in \SS(\R)$ by
\begin{equation*}
[\Hi f](x) = \hbox{$\frac{1}{\pi}$}\Pv \int_\R \frac{f(y)}{x-y}\;\!\d y =
\hbox{$\frac{-i}{\sqrt{2\pi}}$}\int_\R e^{ikx}\;\!\hat{k}\;\!\hat{f}(k)\;\!\d k\
\end{equation*}
and can be continuously extended to a bounded operator in $\H$, still denoted by $\Hi$.
In the previous expression, $\Pv$ denotes the principal value. We also define $\sigma: \H\to \H$ by $[\sigma f](x) =
\hat x f(x)$ for any $f \in \H$ and $x\in \R^*$. Clearly $\sigma$ yields an isomorphism between
$\H_\e$ and $\H_\o$, the subspaces of even, respectively odd, functions of $\H$.

\begin{lemma}\label{lemT}
On $\SS(\R)$, the equality $2\T = i\sigma \Hi +1 $ holds, and thus $\T$ extends continuously to a bounded operator in $\H$, still denoted by $\T$.
\end{lemma}
\begin{proof}
Let $f \in \SS(\R)$. For $x>0$ one has
\begin{equation*}
[(i\sigma \Hi + 1)f](x)=\hbox{$\frac{1}{\sqrt{2\pi}}$}
\int_\R e^{ikx} (\hat x \hat k+1) \hat f(k)\;\! \d k=
\hbox{$\frac{2}{\sqrt{2\pi}}$}\int_{0}^{+\infty}
e^{ik|x|} \hat f(k\hat{x})\;\!\d k.
\end{equation*}
For $x<0$, one has
\begin{eqnarray*}
[(i\sigma \Hi + 1)f](x)&=&
\hbox{$\frac{1}{\sqrt{2\pi}}$}
\int_\R
e^{ikx} (\hat x \hat k+1) \hat f(k) \;\!\d k=
\hbox{$\frac{2}{\sqrt{2\pi}}$}
\int_{-\infty}^{0} e^{ikx} \hat f(k)\;\! \d k \\
&=&  \hbox{$\frac{2}{\sqrt{2\pi}}$}
\int_{0}^{+\infty} e^{-ikx} \hat f(-k)\;\!\d k =
\hbox{$\frac{2}{\sqrt{2\pi}}$}\int_{0}^{+\infty}
e^{ik|x|} \hat f(k\hat{x})\;\!\d k.
\end{eqnarray*}
The last statement follows then by density.
\end{proof}

\subsection{$\T$ as a function of $A$}
Recall \cite{Jen} that the dilation group is represented on $\H$ by
\begin{equation*}
[U_\tau f](x) = e^{\tau/2} f(e^\tau x)\ ,
\end{equation*}
with $f \in \H$, $\tau \in \R$ and $x \in \R$.
Its self-adjoint generator $A$ is formally given by
$\hbox{$\frac{1}{2i}$}(X\nabla + \nabla X)$,
where $X$ is the position operator and $\nabla = \frac{d}{dx}$.  These
operators are all essentially self-adjoint on $\SS(\R)$.
It is easily observed that the formal equality $\F\;\!A\;\!\F^*=-A$ holds. More precisely, for any essentially bounded function $\varphi$ on $\R$, one has $\F\varphi(A)\F^* = \varphi(-A)$.
Furthermore, since $A$ acts only on the radial coordinate, the operator $\varphi(A)$ leaves $\H_\e$ and $\H_\o$ invariant.
For that reason, we can consider a slightly more complicated operator than $\varphi(A)$. Let $\varphi_\e$, $\varphi_\o$ be two essentially
bounded functions on $\R$. Then $\varphi(A):\H \to \H$
defined on $\H_\e$ by $\varphi_\e(A)$ and on $\H_\o$ by
$\varphi_\o(A)$, is a bounded operator.

We first state a result about the Mellin transform.

\begin{lemma}\label{Jensen}
Let $\varphi$ be an essentially bounded function on $\R$ which is the
image under the Fourier transform of a distribution $\check\varphi$ on
$\R$ with rapid decay to $0$ at infinity.
Then, for any $f \in \SS(\R)$ with compact support and any $x \in \R^*$ one has
\begin{equation}\label{eq-Jensen}
[\varphi(A)f](x) =
\hbox{$\frac{1}{\sqrt{2\pi}}$} \int_0^\infty\check{\varphi}
\big(\ln(\hbox{$\frac{|x|}{y}$})\big)\;\!\big(\hbox{$\frac{|x|}{y}$}\big)^{1/2}\;\!f(y \hat{x})\;\! \hbox{$\frac{\d y}{|x|}$}\ ,
\end{equation}
where the r.h.s.~has to be understood in the sense of distributions.
\end{lemma}

\begin{proof}
The proof is a simple application for $n=1$ of the general formulas
developed in \cite[p.~439]{Jen}.
Let us however mention that the convention of this reference on the
minus sign for the operator $A$ in its spectral representation
has not been followed.
\end{proof}

\begin{lemma}\label{TetR}
The equality
$i\sigma \Hi = -R(A)$ holds in $\H$, where $R(A)$ is given in \eqref{eq-R}.
\end{lemma}

\begin{proof}
For any $\epsilon>0$, let us define the integral kernel
\begin{equation*}
I_\epsilon(x,y) = \hbox{$\frac{1}{2\pi}$}\int_\R
\; e^{-ik(x-y)}\;\! \hat{y}\;\! \hat{k} \;\! e^{-\epsilon|k|} \;\!\d k\ ,
\end{equation*}
for $x,y\in\R^*$. We denote by $I_\epsilon$ the operator it defines on $\H$.
An application of the theorems of Fubini and Lebesgues shows that for any $f \in \SS(\R\setminus\{0\})$:
\begin{equation*}
\lim_{\epsilon\to 0} \;[I_\epsilon f](x) = [\F\; (i\sigma \Hi) \;
\F^* f](x)\ .
\end{equation*}
On the other hand, one easily obtains :
$I_\epsilon(x,y) = - \frac{2i\hat{y}(x-y)}{2\pi[(x-y)^2+\epsilon^2]}$.
Now, let us assume additionally that $f$ is an even function, and let $x>0$. Then one has
\begin{equation*}
[I_\epsilon f](x) = - \hbox{$\frac{2i}{2\pi}$}\int_0^\infty
\Big[\frac{x-y}{(x-y)^2+\epsilon^2} -
  \frac{x+y}{(x+y)^2+\epsilon^2}\Big] f(y) \;\d y.
\end{equation*}
Comparison with \eqref{eq-Jensen} yields therefore that
$ \lim_{\epsilon\to 0} I_\epsilon f = \varphi_e(A) f$, with
\begin{equation*}
\check\varphi_e\big(\ln(\hbox{$\frac{x}{y}$})\big) =  -\hbox{$\frac{i}{\sqrt{2\pi}}$}
\Big[\Pv\Big(\frac{1}{\sinh\big(\frac{1}{2}\ln(\frac{x}{y})\big)}\Big) -
\frac{1}{\cosh\big(\frac{1}{2}\ln(\frac{x}{y})\big)} \Big].
\end{equation*}
Similarly, for $f$ odd and $x>0$, one obtains
\begin{equation*}
[I_\epsilon f](x) = - \hbox{$\frac{2i}{2\pi}$}\int_0^\infty
\Big[\frac{x-y}{(x-y)^2+\epsilon^2} +
\frac{x+y}{(x+y)^2+\epsilon^2}\Big] f(y) \;\d y.
\end{equation*}
and $\lim_{\epsilon\to 0} I_\epsilon f = \varphi_o(A) f$ with
\begin{equation*}
\check\varphi_o\big(\ln(\hbox{$\frac{x}{y}$})\big) =  -\hbox{$\frac{i}{\sqrt{2\pi}}$}
\Big[\Pv\Big(\frac{1}{\sinh\big(\frac{1}{2}\ln(\frac{x}{y})\big)}\Big) +
\frac{1}{\cosh\big(\frac{1}{2}\ln(\frac{x}{y})\big)} \Big].
\end{equation*}
Using that the Fourier transform of
$\Pv\Big(\frac{1}{\sinh(\frac{\cdot}{2})}\Big)$ is $-i\sqrt{2\pi}\tanh(\pi \cdot)$ and
that of $\frac{1}{\cosh(\frac{\cdot}{2})}$ is $ \frac{\sqrt{2\pi}}{\cosh(\pi \cdot)}$,
one obtains explicit expressions for $\varphi_e$ and $\varphi_o$.
By density one finally gets that
$$i\sigma \Hi = \varphi_e(-A)\;\!P_e + \varphi_o(-A)P_o,$$
and then the statement follows from the equalities
$\hbox{$\frac{1}{2}$}\big(1-R(A)\big)\equiv \T = \hbox{$\frac{1}{2}$}(i\sigma \Hi +1)$.
\end{proof}
We have thus verified equations \eqref{newformula} and \eqref{eq-R} provided we
show compactness of $K$. But before doing that we wish to point out
the remarkable fact that
for one-dimensional point interactions the formulae obtained above are correct with
$K=0$ \cite{KR1}. In fact, for the $\delta$-interaction of strength
$\alpha\in \R\cup\{\infty\}$ --
the parameter $\alpha$
describes the boundary condition of the wave function
$\Psi'(0_+)-\Psi'(0_-) = \alpha \Psi(0)$ which can be formally
interpreted as arising from a potential $V=\alpha\delta$ where
$\delta$ is the Dirac $\delta$-function at $0$ -- the wave operator is
given by
\begin{equation} \label{eq-wave-point}
\Omega^\alpha_- =
1+\hbox{$\frac{1}{2}$}
\Big[1+\tanh(\pi A)  + i \big(\cosh(\pi A)\big)^{-1}\Big] \;\left(\frac{2\sqrt{-\Delta}-i\alpha}{2\sqrt{-\Delta}+i\alpha}
-1\right) P_e\ .
\end{equation}

\subsection{Compactness of $K$}

Let us first observe that the expression for $K(x,k)$ can be simplified. Indeed, it is easily observed that for $x\geq 0$ one has
\begin{equation*}
K(x,k)= \int_x^\infty \hbox{$\frac{\sin(|k|\;\!(y-x))}{|k|}$}\;\! V(y)\;\! \Psi(y,k)\;\!\d y\
\end{equation*}
and for $x<0$ one has
\begin{equation*}
K(x,k)= \int^x_{-\infty} \hbox{$\frac{\sin(|k|\;\!(x-y))}{|k|}$} \;\! V(y)\;\! \Psi(y,k)\;\!\d y
\end{equation*}
We suspect that the corresponding integral operator $K$ is compact for any potential $V$ belonging to $L^1_1(\R)$. However, we give below a simple proof that under a stronger assumption on $V$ the integral operator $K$ is even Hilbert-Schmidt.

\begin{proposition}
If $V$ belongs to $L^1_\rho(\R)$ for some $\rho > 5/2$, then the operator $K$ is Hilbert-Schmidt.
\end{proposition}

\begin{proof}
The proof consists in showing that the map
$\R\times \R \ni (x,k) \mapsto K(x,k) \in \C$ belongs to $L^2(\R\times \R)$. Since the above definition for $K(\cdot,\cdot)$ is symmetric for $x\in \R_\pm$, we shall concentrate only on the case $x\geq 0$, the case $x<0$ being analogous.

a) Let us first consider the case $|k| \geq 1$. It is known that the function $\Psi(\cdot,\cdot)$ is bounded, independently of $x \in \R$ and $k \in \R\setminus(-1,1)$.  Indeed, for $k>0$, one has
$\Psi(x,\pm k) = e^{\pm ixk}\;\!T(k)\;\!m_\pm(x,k)$, where $T(\cdot)$ is the transmission coefficient and $m_\pm$ are the Jost functions. Furthermore, it is known that $|T(k)|\leq 1$  \cite[Thm.~1]{DT} and that $|m_\pm(x,k)|$ is bounded, independently of $x \in \R$ and $k \in \R\setminus(-1,1)$ \cite[Lem.~1]{DT}.  Thus, it follows that
\begin{eqnarray*}
|K(x,k)|  & = &  \Big| \int_0^\infty \hbox{$\frac{\sin(|k|\;\!y)}{|k|}$}\;\! V(x+y)\;\! \Psi(x+y,k)\;\!\d y \Big| \\
& \leq & \hbox{$\frac{c}{|k|}$} \int_0^\infty \langle x+y\rangle^{-1}\;\!\langle x+y \rangle^1\;\! |V(x+y)| \;\! \d y \\
& \leq & \hbox{$\frac{c}{|k|}$}\;\! \langle x\rangle^{-1}
\int_\R \langle y \rangle\;\! |V(y)|\;\!\d y \\
&\leq & \hbox{$\frac{d}{|k|}$}\;\! \langle x\rangle^{-1}\ ,
\end{eqnarray*}
where $c$ and $d$ are two constants independent of $x$ and $k$. Then one clearly has that $|K(x,k)| \in L^2\big(\R_+\times \R\setminus (-1,1)\big)$.

b) Let us now assume that $k \in (0,1)$.  By taking into account the bound $\frac{|\sin(|k|y)|}{|k|}\leq \frac{4y}{1+|k|y}$, one has for any $\alpha >0$:
\begin{eqnarray*}
|K(x,\pm k)| & \leq & \int_0^\infty \hbox{$\frac{4 y}{1+|k|y}$} \;\! |V(x+y)| \;\!|\Psi(x+y,\pm k)| \;\!\d y \\
& \leq & 4 \langle x \rangle^{-\alpha}\int_0^\infty \langle x+y\rangle^{1+\alpha}\;\!|V(x+y)|\;\!\big|e^{\pm ik(x+y)}\;\! T(k)\;\!m_\pm (x+y,k)\big|\;\! \d y \\
& \leq & 4 \langle x \rangle^{-\alpha}\int_0^\infty \langle x+y\rangle^{2+\alpha}\;\!|V(x+y)|\;\!\Big[ \langle x+y\rangle^{-1}\big(\big|m_\pm(x+y, k)-1\big|+1\big)\Big]\;\!\d y\ .
\end{eqnarray*}
Since $|m_\pm(x,k)-1|$ grows at most linearly in $|x|$, independently of $k$ \cite[Lem.~1]{DT}, the term into square brackets is bounded, independently of $x,y$ and $k$. Thus, one easily obtains that $|K(x,k)|$ belongs to $L^2\big(\R_+\times (-1,1)\big)$ if $\alpha>1/2$, that is if $\rho>5/2$.
\end{proof}

\begin{theorem}\label{cor-newformula}
If $V$ belongs to $L^1_\rho(\R)$ for some $\rho > 5/2$, then
formula \eqref{newformula} holds with $K$ compact.
\end{theorem}
We note again that the first factor $ \big(1 - R(A)\big)$ is universal
in the sense that it does not depend on the potential.
Since $\Omega_+ = \Omega_- S^*$
the analogous formula for the other wave operator reads
\begin{equation}\label{omegaplus}
\Omega_+ = 1+ \hbox{$\frac{1}{2}$} \big(1 + R(A)\big)\big(S(-\Delta)^* -
1\big) + K'
\end{equation}
with $K'=KS^*$ as well compact.

\section{Levinson's theorem as an index theorem}

The Introduction containing a brief description of our topological approach
of Levinson's theorem, we directly start by defining the $C^*$-algebras.
The algebra $\E$ is constructed with the help of the generator $A$ of
dilations and of the operator $B:=\frac{1}{2}\ln(H_0)$ defined by
functional calculus. The crucial property is that
$A$ and $B$ satisfy the canonical commutation
relation $[A,B]=i$ so that $A$ generates translations in $B$ and vice versa,
\begin{equation}\label{eq-cr}
e^{iBt} A e^{-iBt} = A+t, \quad e^{iAs} B e^{-iAs} = B-s.
\end{equation}
Furthermore, both operators leave the subspaces $\H_e$ and $\H_o$ invariant. More precisely, for any essentially bounded functions $\varphi$ and $\eta$ on $\R$, the operator $\varphi(A)\eta(B)$ leaves both subspaces invariant. For that reason, we shall subsequently identify the Hilbert space $\H=\H_e\oplus \H_o$ with $L^2(\R_+,\C^2)\equiv L^2(\R_+)\otimes \C^2$. And more generally, we can consider functions $\varphi, \eta$ defined on $\R$ and taking values in $M_2(\C)$.

Now, let $\E$ be the closure in $\B(\H)$ of the algebra generated by elements of the form $\varphi(A)\psi(H_0)$, where $\varphi$ is a
continuous function on $\R$ with values in $M_2(\C)$ which converges at $\pm \infty$, and $\psi$ is a continuous function $\R_+$ with values in $M_2(\C)$ which converges at $0$ and
at $+\infty$. Stated differently, $\varphi\in C\big(\bR,M_2(\C)\big)$, where $\bR=[-\infty,+\infty]$, and $\psi \in C\big(\bRp,M_2(\C)\big)$ with
$\bRp=[0,+\infty]$. Let $\J$ be the norm closed
algebra generated by $\varphi(A)\psi(H_0)$ with functions $\varphi$ and $\psi$ for which the above limits vanish. Obviously, $\J$ is an ideal in $\E$, and the same algebras are obtained if $\psi(H_0)$ is replaced by $\eta(B)$ with $\eta \in C\big(\bR,M_2(\C)\big)$ or
$\eta \in C_0\big(\R,M_2(\C)\big)$, respectively.

In a completely different context, these algebras have already been  studied in \cite{Georgescu}. The authors introduced them in terms of the operator $X$ and $-i\nabla$ on $L^2(\R,E)$, with $E$ an auxiliary Hilbert space, possibly of infinite dimension. In that situation, the corresponding functions $\varphi$ and $\eta$ are norm continuous function on $\bR$ with values in $\K(E)$. The isomorphism between our algebras and the algebras introduced in \cite[Sec.~3.5]{Georgescu} is given by the Mellin transform. Indeed, $A$ is unitarily equivalent through this transform to the operator $X$ on $L^2\big(\R,M_2(\C)\big)$, and the operator $B$ is equal to $-i\nabla$ in this representation.
For that reason, we shall freely use the results obtained in that reference, and refer to it for the proofs. In particular, it is proved that $\J$ is equal to $\K(\H)$, and an explicit description of the quotient $\E/\J$ is given, which we specify now in our context.

To describe the quotient $\E/\J$ we consider the square
$\square:=\bRp\times \bR$ whose boundary $\partial \square$ is the
union of four parts: $\partial \square =B_1\cup B_2\cup B_3\cup
B_4$, with $B_1 = \{0\}\times \bR$, $B_2 = \bRp \times \{+\infty\}$,
$B_3 = \{+\infty\}\times \bR$ and $B_4 = \bRp\times \{-\infty\}$. We
can then view $C\big(\partial \square,M_2(\C)\big)$ as the subalgebra
of
\begin{equation*}
C\big(\bR,M_2(\C)\big)\oplus C\big(\bRp,M_2(\C)\big)
\oplus C\big(\bR,M_2(\C)\big)\oplus C\big(\bRp,M_2(\C)\big)
\end{equation*}
given by elements
$(\Gamma_1,\Gamma_2,\Gamma_3,\Gamma_4)$ which coincide at the
corresponding end points, that is,
$\Gamma_1(+\infty)=\Gamma_2(0)$,
$\Gamma_2(+\infty)=\Gamma_3(+\infty)$,
$\Gamma_3(-\infty)=\Gamma_4(+\infty)$,
$\Gamma_4(0)=\Gamma_1(-\infty)$.
The following lemma corresponds to results obtained in
\cite[Sec.~3.5]{Georgescu} rewritten in our framework.

\begin{lemma}\label{image}
$\E/\J$ is isomorphic to
$C\big(\partial\square,M_2(\C)\big)$. Furthermore, for any   $\varphi\in C\big(\bR,M_2(\C)\big)$ and $\psi \in C\big(\bRp,M_2(\C)\big)$, the image of $\varphi(A)\psi(H_o)$ through the quotient map $q: \E \to C\big(\partial\square,M_2(\C)\big)$  is given by $\Gamma_1(A) = \varphi(A)\psi(0)$, $\Gamma_{2}(H_0) =
\varphi(+\infty)\psi(H_0)$,
$\Gamma_{3}(A) = \varphi(A)\psi(+\infty)$ and $\Gamma_{4}(H_0) =
\varphi(-\infty)\psi(H_0)$.
\end{lemma}

The following proposition contains the proof of the equality of the
two topological invariants mentioned in the Introduction. In fact,
this equality could be borrowed from a more general result of
\cite{BC}, but we prefer to give a short and more fashionable
proof. Recall that by Atkinson's theorem the image $q(F)$ of
any Fredholm operator $F\in\E$ in the algebra
$C\big(\partial\square,M_2(\C)\big)$ is invertible. We define the
winding number $w\big(q(F)\big)$ to be the winding number of
$\partial\square \ni z \mapsto \det [q(F)(z)]\in \C^*$ with orientation
of $\partial\square$ chosen right around in Figure~1.
\begin{proposition}
For any Fredholm operator $F$ in $\E$,
the winding number $w\big(q(F)\big)$ satisfies  the equality
\begin{equation*}
w\big(q(F)\big) = \index(F)\ .
\end{equation*}
\end{proposition}

\begin{proof}
For simplicity, we shall write $\CC$ for $C\big(\partial\square,M_2(\C)\big)$.
Let us first consider the short exact sequence
\begin{equation*}
0\to \J\to \E\stackrel{q}{\to} \CC \to 0\ .
\end{equation*}
Since $\J=\K(\H)$, one has $K_1(\J)=0$, and the six-term exact
sequence in $K$-theory \cite{Roerdam} associated with the above
sequence reads
\begin{equation}\label{6t}
0 \to K_1(\E)\to K_1(\CC) \stackrel{\rm
ind}{\longrightarrow} K_0(\J) \to K_0(\E)\to K_0(\CC) \to 0\, .
\end{equation}
It is well known that $K_0(\J)\cong \Z$, with a morphism given by the
trace $\Tr$ on $\H$. Furthermore, $K_1(\CC) \cong
K_1\big(C(\S),M_2(\C)\big)\cong\Z$ with a morphism given by the
winding number of the pointwise determinant of the $2 \times 2$
matrix, simply denoted by $w$. Thus, the sequence \eqref{6t} becomes
\begin{equation*}
0 \to K_1(\E)\to \Z \stackrel{n\;\!  \mbox{\rm \footnotesize
id}}{\longrightarrow} \Z \to K_0(\E)\to K_0(\CC) \to 0\,
\end{equation*}
where $n$ is some integer which can be obtained from the equation
$\Tr\big(\ind([u]_1)\big) = n\;\!w([u]_1)$
evaluated on any particular non-trivial element.
More precisely, one has $\ind([u]_1) = [1-W^* W]_0-[1-W W^*]_0$ \cite[Prop.~9.2.4]{Roerdam}, and hence $\Tr\big(\ind([u]_1)\big) = \index(W)$, provided that $W$ is a lift of $u$ which is a partial isometry.
As a particular element of $\E$, let us consider $W:=\Omega_-^{\alpha}$, the wave operator for the
$\delta$-interaction at strength $\alpha<0$ recalled in \eqref{eq-wave-point} and obtained in \cite{KR1}.
By an explicit calculation performed in this reference, one shows that $w(q(\Omega_-^{\alpha}))=-1$. Since the corresponding operator $H$ describing the point interaction has precisely one eigenvalue, we have $\index(\Omega_-^{\alpha})=-1$, and hence $n=1$.
\end{proof}

\begin{theorem}\label{mainthm}
If $V$ belongs to $L^1_\rho(\R)$ for some $\rho > 5/2$, both operators $\Omega_\pm$ belong to $\E$. Furthermore, the following equality holds:
\begin{equation}\label{nono}
w\big(q(\Om_\pm)\big) = -\mbox{\em Tr}(P_p)\ .
\end{equation}
\end{theorem}
\begin{proof}
That $\Omega_\pm\in\E$ follows directly from the description of these
operators given in Theorem \ref{cor-newformula} and in the remark
following it. The equality follows from the previous proposition.
\end{proof}

\subsection{Decomposition of the winding number}\label{secwin}

The four parts given by the image of the wave operator
$q(\Om) = (\Gamma_1,\Gamma_2,\Gamma_3,\Gamma_4)$ are easily
determined with the help of Lemma \ref{image}. Indeed, since $R(-\infty) = 1$ one has
$$\Gamma_4(H_0) = 1.$$
From $R(+\infty) = -1$, it follows that
$$\Gamma_2(H_0) = 1 + \hbox{$\frac{1}{2}$}\big(1-R(+\infty)\big)(S(H_0)-1) = S(H_0).$$
Finally, one has
\begin{equation}\label{gamma1}
\Gamma_1(A) = 1 + \hbox{$\frac{1}{2}$}\big(1-R(A)\big)(S(0)-1),
\end{equation}
and
$$\Gamma_3(A) = 1 + \hbox{$\frac{1}{2}$}\big(1-R(A)\big)(S(+\infty)-1).$$
To better visualise these results, one may consider the  following figure:
\begin{center}
\epsfysize=5cm
\epsfbox{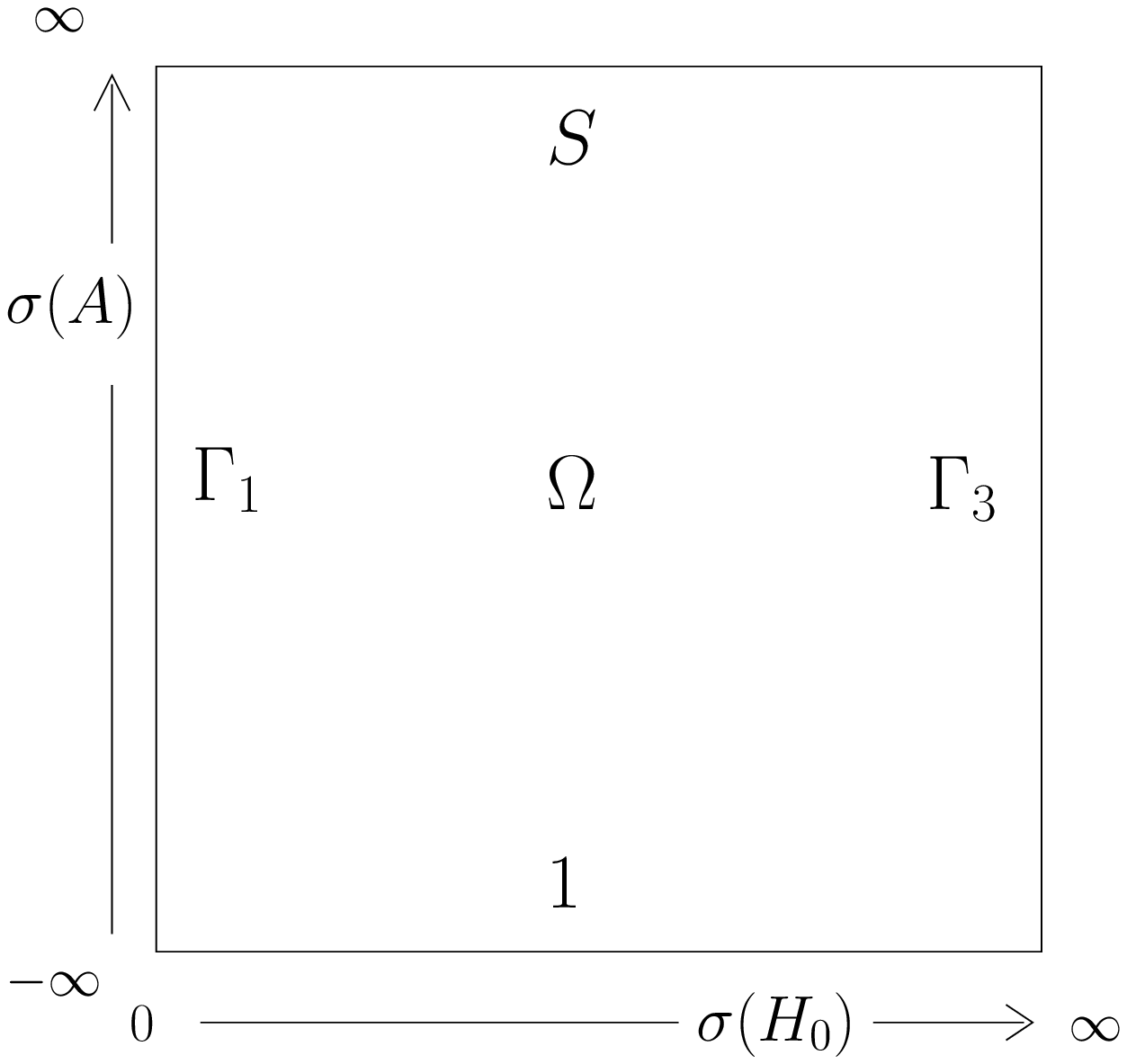}
\end{center}
\begin{itemize}
\item[]
Figure 1: The square
$\square=\overline{\sigma(H_0)}\times\overline{\sigma(A)}$ with the
wave operator $\Omega$ and the limits $\Gamma_j$, $j=1,\ldots,4$ resulting from applying the quotient map $q$. One always has
$\Gamma_2=S$ and $\Gamma_4=1$ whereas $\Gamma_1$ and $\Gamma_3$ depend
on the value of the $S$ matrix at $0$ and $\infty$, respectively.
\end{itemize}
Consequently, the winding number $w\big(q(\Om)\big)$ is the sum of four terms, each side
of the square contributing for one. If the functions $\Gamma_j$ are
piecewise differentiable and if the following integrals exist, we can use the trace formula for determining the winding number, namely
\begin{equation}\label{tracef}
w\big(q(\Om)\big) =\sum_{j=1}^4 w_j,\quad
w_j := \hbox{$\frac{1}{2\pi i}$} \int_{B_j} \tr[\Gamma_j^{-1}\;\!\d \Gamma_j]\ .
\end{equation}
We find immediately that $w_4 = 0$ and that
$$w_2 = \hbox{$\frac{1}{2\pi i}$} \int_0^\infty
\tr[S^*(\lambda) S'(\lambda)]\d \lambda$$
is (minus) the integral of the trace of the time delay.
This expression exhibits our choice of orientation in the calculation
of the winding number, namely it goes from energy $0$ to energy
$\infty$ along the side $B_2$ of the square.
 Comparing \eqref{eq-lev1} with \eqref{eq-lev21} we see therefore that
 the correction term $\nu$ arises now on the l.h.s.\ of the equality
 from the possible contribution of $\Gamma_1$ and $\Gamma_3$ to the
winding number. Whereas for point interactions, $w_3$ need not to be
$0$ \cite{KR1}, the known fact that for potential scattering
$S(+\infty) = 1$ implies that $\Gamma_3(A)=1$ and $w_3=0$.

We now determine the contribution coming from $w_1$. For that we could
use the known results of the literature about the form of $S(0)$ but
we will give an independent argument. Its only ingredients are the
unitarity of $\Gamma_1(A)$, $S(0)$ and $R(A)$, which follows from the
fact that $\Omega$ is a isometry, and the explicit form of $R(A)$.
More precisely, in the decomposition of $\H$ into $\H_e \oplus \H_o$, the operator $R(A)$ takes the form
$\left(\begin{smallmatrix}
r_e(A) & 0 \\
0 & r_o(A)
\end{smallmatrix}\right)$, with
\begin{equation*}
r_{e/o}(x)=-\tanh(\pi x) \mp i\cosh^{-1}(\pi x) \qquad \forall \; x \in \R\ .
\end{equation*}

\begin{proposition}
Either $\det\big(S(0)\big)=-1$ and then $S(0)=\pm \left(\begin{smallmatrix}
-1 & 0 \\
0 & 1
\end{smallmatrix}\right)$, or $\det\big(S(0)\big)=1$ and then
$S(0)= \left(\begin{smallmatrix}
a & b \\
-\bar{b} & a
\end{smallmatrix}\right)$
with $a\in\R$, $b\in\C$, and $a^2+|b|^2=1$.
Moreover, $w_1=\mp\frac{1}{2}$ in the first case, and $w_1=0$ in the second one.
\end{proposition}

\begin{proof} Let us set $S:=S(0)$ and $R:=R(A)$.
By rewriting $\Gamma_1(A)$ as $\frac{1}{2}\big[(1-R)S + (1+R)\big]$, the condition $\Gamma_1(A)\;\!\Gamma^*_1(A)=1$ yields
\begin{equation*}
 4 = (1-R)(1-R^*) + (1+R)(1+R^*) + (1-R)S(1+R^*) + (1+R)S^*(1-R^*)
\end{equation*}
which implies that $ (1-R)S(1+R^*) + (1+R)S^*(1-R^*) = 0$. By
multiplying both sides of this equality with $R$ one obtains
$ (1-R)S(R+1) + (1+R)S^*(R-1) = 0$, or equivalently
\begin{equation*}
X - RXR + [Y,R] = 0
\end{equation*}
where $X=S-S^*$ and $Y=S+S^*$.
In the basis mentioned above in which $R$ is diagonal, the previous equality is equivalent to
$$ \left(\begin{array}{cc}
(1-r_e^2)X_{ee} &(r_o-r_e)Y_{eo} \\
 (r_e-r_o)Y_{oe} & (1-r_o^2)X_{oo}
\end{array}\right) = 0$$
in which the equality  $r_er_0 = 1$ has been taken into account.
This equality implies that $S_{ee}$ and $S_{oo}$ are real, and that and $S+S^*$ is diagonal. This fact together with the unitarity of $S$ imply that
$\det(S)=\pm 1$ and that the matrix $S$ has a form as stated.

For the calculation of $w_1$, it follows from \eqref{tracef} that :
\begin{eqnarray*}
w_1 &=& \hbox{$\frac{1}{2\pi i}$} \int_{B_1} \tr \Big[\hbox{$\frac{1}{4}$}[(1+R^*)+S^*(1-R^*)]\;\!(\d R)\;\!(1-S)\Big]\\
&=& \hbox{$\frac{1}{8\pi i}$}\int_{B_1} \tr\big[(2-S-S^*)R^*\d R\big]\\
  &=& \hbox{$\frac{1}{4}$}[-(1-S_{ee}) + (1-S_{oo})] \\
&=&\hbox{$\frac{1}{4}$}(S_{ee}-S_{oo})\ .
\end{eqnarray*}
where we have used that $\int_{B_1} \tr\big[(S-S^*)\d R\big] = 0$, and that $\frac{1}{2\pi i}\int_{-\infty}^{\infty} r_e^*\d r_e =
-\frac{1}{2\pi i}\int_{-\infty}^\infty r_o^*\d r_o = -\frac{1}{2}$.
\end{proof}

It is interesting to note that the result on the restriction of the
form for $S(0)$ coincides almost with the possible forms that can
occur in potential scattering, see {\it e.g.}~\cite{AK}.
More precisely, only the case  $S(0)=\left(\begin{smallmatrix}
1 & 0 \\
0 & -1
\end{smallmatrix}\right)$ cannot occur.
It is found that $\det\big(S(0)\big)=-1$ if $H$ does not
admit a resonance at energy zero. This is referred to as the
generic case (g.c.). In this case we thus have $w_1=-\frac{1}{2}$.
The  so-called exceptional case (e.c.)
corresponds to $\det\big(S(0)\big)=1$ and
occurs when such a zero energy resonance exists.

Thus, taking into account that $\Gamma_3 = \Gamma_4=1$ one obtains from \eqref{eq-lev21}
\begin{equation}\label{presquefini}
\hbox{$\frac{1}{2\pi}$}\int_{\R_+} \tr
[i S^*(\lambda)S'(\lambda)]  \d \lambda =
\left\{\begin{array}{ll}
{N -\frac{1}{2}}, & \hbox{g.c.} \\
{N}, & \hbox{e.c.}
\end{array}\right.
\end{equation}
In particular, the correction term $-\nu$ corresponds to $w_1$.
This result \eqref{presquefini} is in accordance with the literature \cite{BGK,BGW,DMa,Ma,Sassoli}.

\begin{remark}\label{rem1}
{\rm
We wish to make clear that our result, namely that Levinsons' theorem
is an index theorem, is quite different from a result
encountered in super\-symmetric quantum mechanics about the
topological stability of the so-called Witten index and its relation
to half-bound states, anomalies and $0$-energy eigenstates.

In super\-symmetric quantum mechanics one is led to consider a
closed densely defined operator $A$ on some Hilbert space and to
compare the dynamics of $H_+=AA^*$ with $H_-=A^*A$.
The partial isometry
arising from the polar decomposition of $A$ yields a unitary
equivalence between the orthogonal complements of $\ker H_-$ and
$\ker H_+$.
The quantity
$ W(A): = \lim_{\beta\to +\infty} \Tr(e^{-\beta H_+} - e^{-\beta
  H_-})$ is a measure to which extend this symmetry
fails to hold between $\ker H_-$ and $\ker H_+$.
$W(A)$ was
introduced by Witten, and is now called Witten index,
who was motivated by supersymmetric quantum
field theory which leads one to consider the differential operator
$A=-i\nabla -i\varphi(X)$ on $\R$ for some mesurable bounded function
$\varphi$ which has finite limits at $\pm \infty$ (see the review
\cite{NiSe} and references therein). As was realized in
\cite{BoBl}
the Witten index for the above model is given by
$W(A) = \frac{1}{2\pi}\int_{\lambda_0}^\infty
\tr \big[i S^*(\lambda) S'(\lambda)\big] \d \lambda$,
where $S=S(H_+,H_-)$ is the scattering operator for the pair
$H_+=-\Delta + \varphi^2(x)+\varphi'(x),H_-=-\Delta +
\varphi^2(x)-\varphi'(x)$, and
$\lambda_0$ is the bottom of the essential spectrum of
$H_-$.
Hence in this model $W(A)$ is a quantity which corresponds to our
$w_2$. But note that the contributions from the non-zero
eigenvalues of $H_+$ and $H_-$ will always cancel in the above comparison.

It is remarkable that $W(A)$
depends only on the limit values of $\varphi$ at $\pm\infty$
\cite{BoBl,BGGSS}.
This topological stability, which was thoroughly analyzed in
\cite{GS} for more general $A$,
is due to the (super)symmetry of the
pair of operators under consideration:
Any slight change in $\varphi$ would affect both operators $H_+$ and
$H_-$.
Another remarkable fact is $W(A)$ need not be integer if $A$ is not a
Fredholm operator. This has drawn a lot of attention and was readily
compared with the phenomenon of anomalies, half-bound states and
corrections they
incite for the analog of Levinson's theorem,
see {\it e.g.}~\cite{BoBl,BGGSS,GS,NiSe}.
But none of the above supersymmetric approach is aimed at
showing that Levinson's theorem is an index theorem, nor does the latter
follow for instance from the topological stability of the
Witten index.}
\end{remark}

\section{Restricted norm convergence}\label{sec5}

In \cite{Georgescu}, an alternative description of the algebra $\E$ in terms of evolution groups is also given. By rephrasing it in our framework, this leads to new propagation estimates. For the time being, these estimates are a corollary of Theorem \ref{mainthm}, but it could also be interesting to obtain them from a direct computation.

We now introduce this new description of $\E$ inspired from \cite{Georgescu}. We also use the convention of that reference, that is: if a symbol like $T^{(*)}$ appears in a relation, it means that this relation holds for $T$ and for its adjoint $T^*$. The function $\chi$ denotes the characteristic function.

\begin{lemma}\label{lemgeo}
A bounded operator $F$ in $\H$ belongs to $\E$ if and only if there exist $\Gamma_1,\Gamma_3 \in C\big(\bR,M_2(\C)\big)$ and $\Gamma_2,\Gamma_4 \in C\big(\bRp,M_2(\C)\big)$ such that the following conditions are satisfied:
\begin{eqnarray*}
\lim_{\varepsilon \to 0}\|\chi(H_0\leq \varepsilon) \;\!\big(F -\Gamma_1(A)\big)^{(*)}\|=0, \quad &&
\lim_{\varepsilon \to +\infty}\|\chi(H_0\geq \varepsilon) \;\!\big(F -\Gamma_3(A)\big)^{(*)}\|=0, \\
\lim_{t \to -\infty}\|\chi(A \leq t) \;\!\big(F -\Gamma_4(H_0)\big)^{(*)}\|=0, \quad &&
\lim_{t \to +\infty}\|\chi(A \geq t) \;\!\big(F-\Gamma_2(H_0)\big)^{(*)}\|=0.
\end{eqnarray*}
Moreover, $F$ belongs to the ideal $\K(\H)$ if the above conditions are satisfied with $\Gamma_j=0$ for $j \in \{1,2,3,4\}$.
\end{lemma}

In particular, for $F=\Omega\equiv \Omega_-$, one has:

\begin{corollary}\label{corolOm}
If $V$ belongs to $L^1_\rho(\R)$ for some $\rho > 5/2$, then
\begin{eqnarray*}
\lim_{\varepsilon \to 0}\|\chi(H_0\leq \varepsilon) \;\!\big(\Omega -\Gamma_1(A)\big)^{(*)}\|=0, \quad &&
\lim_{\varepsilon \to +\infty}\|\chi(H_0\geq \varepsilon) \;\!(\Omega -1)^{(*)}\|=0, \\
\lim_{t \to -\infty}\|\chi(A \leq t) \;\!(\Omega -1)^{(*)}\|=0, \quad && \lim_{t \to +\infty}\|\chi(A \geq t) \;\!\big(\Omega-S(H_0)\big)^{(*)}\|=0,
\end{eqnarray*}
where $\Gamma_1$ is the operator defined in \eqref{gamma1}.
\end{corollary}

Let us point out that this corollary is a much more precise version of
the well known result
\begin{equation*}
(\Omega_--1)\;\!\psi(H_0)\;\!\chi(A\leq 0) \in \K(\H)
\end{equation*}
which holds for any continuous function $\psi$ on $\R_+$ that vanishes in a
neighbourhood of $0$ and that is equal to $1$ in a neighbourhood of
$+\infty$, see for example \cite{Enss,Perry} for a proof of such a
result in the three dimensional case and for its use in the proof of
asymptotic completeness. In the framework of Section \ref{secwin},
this result simply says that $\Gamma_3 = \Gamma_4=1$, but does not say
anything about $\Gamma_1$ and $\Gamma_2$.

Now, by using the relation \eqref{eq-cr}, the third and the fourth
conditions of Corollary \ref{corolOm} can easily be rewritten in terms
of the unitary groups generated by $B$. For example, the third
condition is equivalent to
\begin{equation*}
\lim_{t \to -\infty}\|\chi(A \leq 0) \;\!e^{iBt}\;\!(\Omega
-1)^{(*)}\;\!e^{-iBt}\|=0.
\end{equation*}
Furthermore, the invariance principle and the intertwining relation
allow one to simplify the above expression. Indeed, the following
equalities hold:
\begin{equation}\label{inter}
\Omega_\pm \; e^{-i\ln(H_0)t} = e^{-i\ln(H)t}\; \Omega_\pm\ ,
\end{equation}
where $\ln(H)$ is obtained by functional calculus on the positive part
of the spectrum of $H$. Let us also note that
$e^{-i\ln(H)t} = e^{-i\ln(H)t} E_{ac}(H)$, where
$E_{ac}(H)$ denotes the spectral projection on the absolutely continuous part of $H$. Finally, one obtains :

\begin{proposition}\label{propev}
If $V$ belongs to $L^1_\rho(\R)$ for some $\rho > 5/2$, then
\begin{enumerate}
\item  $\lim_{t \to -\infty}\|\chi(A \leq 0)[ e^{i\ln(H_0)t}\;\!e^{-i\ln(H)t}\;\!\Omega_--1]\|=0$,
\item  $\lim_{t \to -\infty}\|\chi(A \leq 0)[ \Omega_-^*\;\!e^{i\ln(H)t}\;\!e^{-i\ln(H_0)t}-1]\|=0$,
\item $\lim_{t \to +\infty}\|\chi(A \geq 0) [e^{i\ln(H_0)t}\;\!e^{-i\ln(H)t}\;\!\Omega_--S]\|=0$,
\item $\lim_{t \to +\infty}\|\chi(A \geq 0) [\Omega_-^* \;\!e^{i\ln(H)t}\;\!e^{-i\ln(H_0)t}-S^*]\|=0$.
\end{enumerate}
\end{proposition}

\begin{proof}
The proof simply consists in rewriting the last two conditions of
Corollary \ref{corolOm}
in terms of the evolution group generated by $B$ and taking  relation
\eqref{inter} into account.
\end{proof}

Let us add some more words on the first and the second condition
of Corollary \ref{corolOm}. By using the equalities $\chi(H_0\leq
\varepsilon) = \chi(B\leq \frac{1}{2}\ln\varepsilon)$ and
$\chi(H_0\geq \varepsilon) = \chi(B\geq \frac{1}{2}\ln\varepsilon)$,
and the relation \eqref{eq-cr}, these conditions can easily be
rewritten in terms of the unitary groups generated by $A$. For
example, the first condition is equivalent to
\begin{equation*}
\lim_{t \to -\infty}\|\chi(H_0\leq 1) \;\!e^{-iAt}\big(\Omega -\Gamma_1(A)\big)^{(*)}\;\!e^{iAt}\|=0\ .
\end{equation*}
Furthermore, it is easily observed that the following equality holds:
\begin{equation*}
e^{-itA}\;\!\Omega(H_0+V,H_0)\;\!e^{itA} = \Omega(H(t),H_0)\ ,
\end{equation*}
where $H(t) = H_0 + e^{-2t}V(e^{-t} \cdot)$. It thus follows that:

\begin{proposition}\label{propdil}
If $V$ belongs to $L^1_\rho(\R)$ for some $\rho > 5/2$, then
\begin{enumerate}
\item $\lim_{t\to -\infty}\|\chi(H_0\leq 1) \;\!\big(\Omega(H(t),H_0) -\Gamma_1(A)\big)^{(*)}\|=0$,
\item $\lim_{t\to +\infty}\|\chi(H_0\geq 1) \;\!\big(\Omega(H(t),H_0) -1\big)^{(*)}\|=0$,
\end{enumerate}
where $\Gamma_1$ is the operator defined in \eqref{gamma1}.
\end{proposition}

\begin{remark}
{\rm
Let us mention that the study of rescaled operators is quite
common. As an example we
recall a similar construction developed in \cite[Chap.~1]{AGHH} and
comment on its relation to our result.
In that reference the family of operators
$H_\lambda(t)=-\Delta + \lambda(e^t)\;\!e^{-2t}\;\!V(e^{-t}\cdot)$
is introduced where $\lambda$ is a real analytic function in a
neighbourhood of the origin and $\lambda(0)=0$.
It is then proved that as $t \to -\infty$
this operator converges in the norm resolvent sense
to a Schr\"odinger operator with a one point interaction at the origin
of strength $\alpha$.
The parameter $\alpha$ is equal to $\lambda'(0)\int_\R V(x)\d x$ and hence
insensitive to the zero-energy properties of $-\Delta + V$.
A study of the corresponding limit of the wave operator
$\Omega\big(H_\lambda(t),H_0\big)$
yields a convergence
to \eqref{eq-wave-point}. Note that the computation of the
part corresponding to the left side $B_1$ of $q(\Omega_-^\alpha)$ yields
$\Gamma_1^\alpha (A) = P_o + R(A)P_e$, except for $\alpha=0$.

To  compare this with our approach, note that $H(t)=H_{\lambda = 1}(t)$.
Since $\lambda = 1$ contradicts the assumption $\lambda(0)=0$ made
in \cite{AGHH} a direct comparison is not possible.
In view of Proposition \ref{propdil} this better ought to be the case
since the explicit form of $\Gamma_1$ given in \eqref{gamma1} shows that
the corresponding limit as $t \to -\infty$ highly depends on the
existence or the absence of a $0$-energy resonance for
$-\Delta + V$, a result much closer in spirit to the corresponding
one obtained in \cite{AGHH} for systems in $\R^3$, where the function
$\lambda=1$ is allowed. We observe that $\Gamma_1 = \Gamma_1^\alpha$
in the generic case and if $\alpha\neq 0$.

We recall that we consider convergence in a norm restricted sense
of the wave operators
(more precisely norm convergence after multiplication
with the projections
$\chi(H_0\leq 1)$ or $\chi(H_0\geq 1)$). The relation of this kind
of convergence with convergence in the norm resolvent sense of
Hamiltonians is not yet established.
A deeper and independent study of the content of Proposition
\ref{propdil} and of its relation
with the known results on rescaled operators would certainly be of interest.}
\end{remark}

\end{document}